\documentclass[10pt, journal, letterpaper]{IEEEtran}

\usepackage{amsmath}
\usepackage{amssymb} %
\usepackage{amsthm}
\usepackage{color,soul}
\usepackage{balance} %
\usepackage[dvipsnames]{xcolor}
\usepackage{siunitx}
\usepackage{graphicx}
\usepackage{subcaption}

\newcommand{\figsp}{{\vspace{-10pt}}}

\begin{document}

\title{Design and Performance Evaluation of 
SEANet, a Software-defined Networking Platform for the Internet of Underwater Things
}

\author{\IEEEauthorblockN{Deniz Unal,
Sara Falleni,
Kerem Enhos,
Emrecan Demirors,
Stefano Basagni,
Tommaso Melodia}\\
\IEEEauthorblockA{Institute for the Wireless Internet of Things, Northeastern University, Boston, MA, U.S.A.\\
E-mail: \{unal.d, falleni.s, enhos.k, e.demirors, s.basagni, t.melodia\}@northeastern.edu}
\thanks{This work was supported by the National Science Foundation under Grant CNS-1763964 and Grant CNS-1726512.}
}

\maketitle
\thispagestyle{empty}

\begin{abstract}
This paper presents the design and performance evaluation of the SEANet platform, a software-defined acoustic modem designed for enhancing underwater networking and Internet of Underwater Things (IoUT) applications. 
Addressing the limitations of traditional acoustic modems, which suffer from low data rates and rigid architectures, SEANet introduces a versatile, adaptive framework capable of reconfiguring all layers of the protocol stack in real-time to accommodate diverse marine applications. 
The platform integrates high-performance, wideband data converters with modular hardware and software components, enabling real-time adaptation to changing environmental and operational conditions. 
Experimental evaluations conducted in oceanic settings demonstrate the SEANet capability to significantly exceed the performance of existing commercial underwater modems, supporting data rates up to 150 kbit/s and effectively doubling the performance metrics of conventional systems. 
Our robust testing also highlights the SEANet proficiency in channel estimation, Orthogonal Frequency-Division Multiplexing (OFDM) link establishment, and interoperability through the JANUS communication standard. 
Our results underscore the SEANet potential to transform underwater communication technologies, providing a scalable and efficient solution that supports high data rate applications and fosters the expansion of IoUT deployments.
\end{abstract}


\section{Introduction}
\label{sec:introduction}

Exploring and monitoring the oceans are increasingly critical for life on earth, as a thorough understanding of marine environments would help with fighting climate change, enabling applications for the \emph{Blue Economy}, and contributing to our livelihood and wellness.
These tasks and applications require robust and sustainable forms of communication for gathering and delivering data from devices deployed underwater.
Particularly, they would require forms of networking similar to those implemented on land by the Internet of Things (IoT), which extend the Internet networking paradigm to allow all sorts of ``things'' to communicate.
This \emph{Internet of Underwater Things (IoUT)} would provide support for ocean data to be collected and delivered to shore, to create the interface between the ocean and the digital world, to enable real-time decision making on ocean data, and ultimately enable applications that are now prohibitively expensive or impossible~\cite{MohsanMOA22}.

\begin{figure}[]
     \centering
     \begin{subfigure}[b]{0.48\columnwidth}
         \centering
         \includegraphics[width=\textwidth]{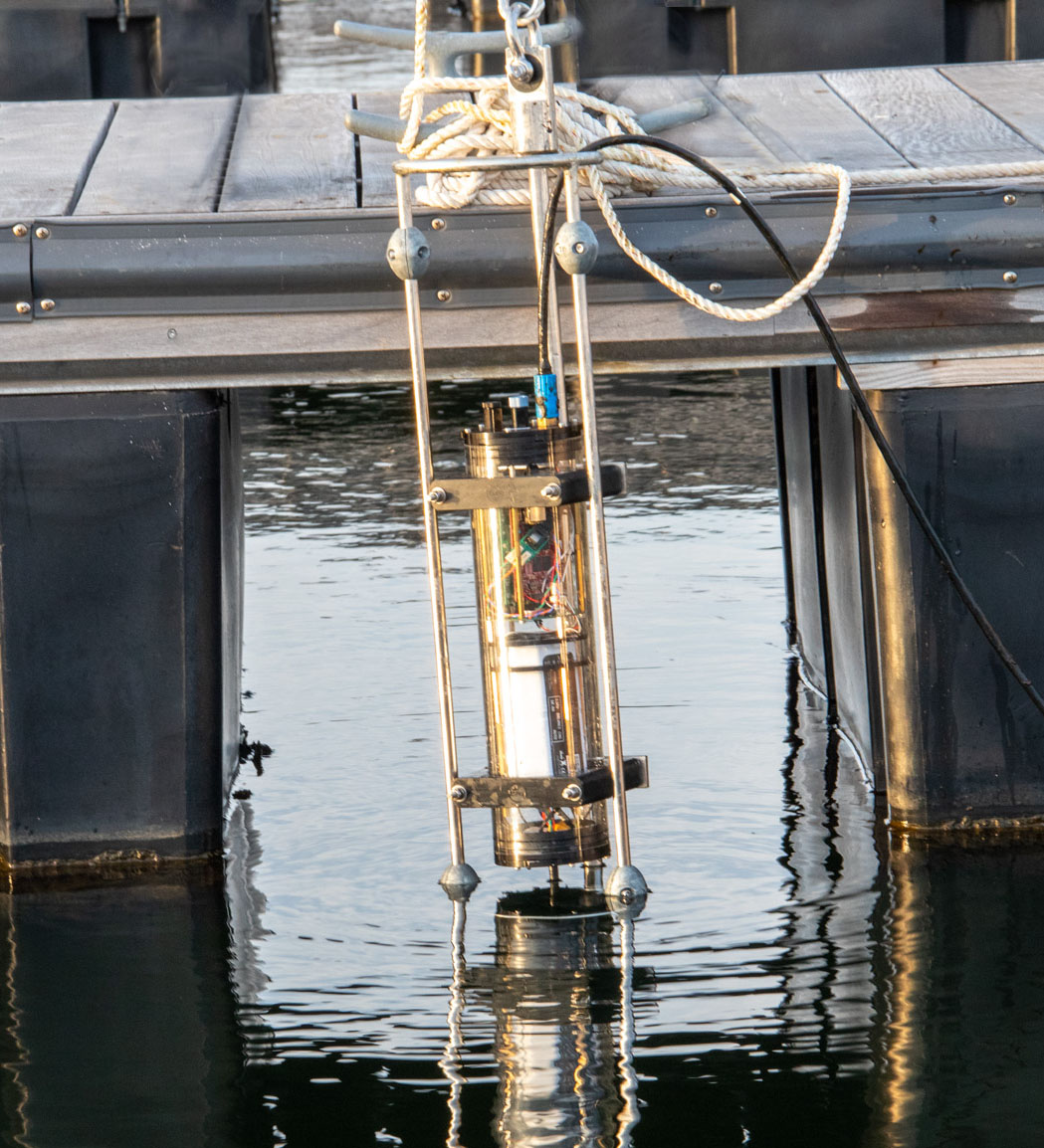}
     \end{subfigure}
     \hspace{-0cm}
     \begin{subfigure}[b]{0.48\columnwidth}
         \centering
         \includegraphics[width=\textwidth]{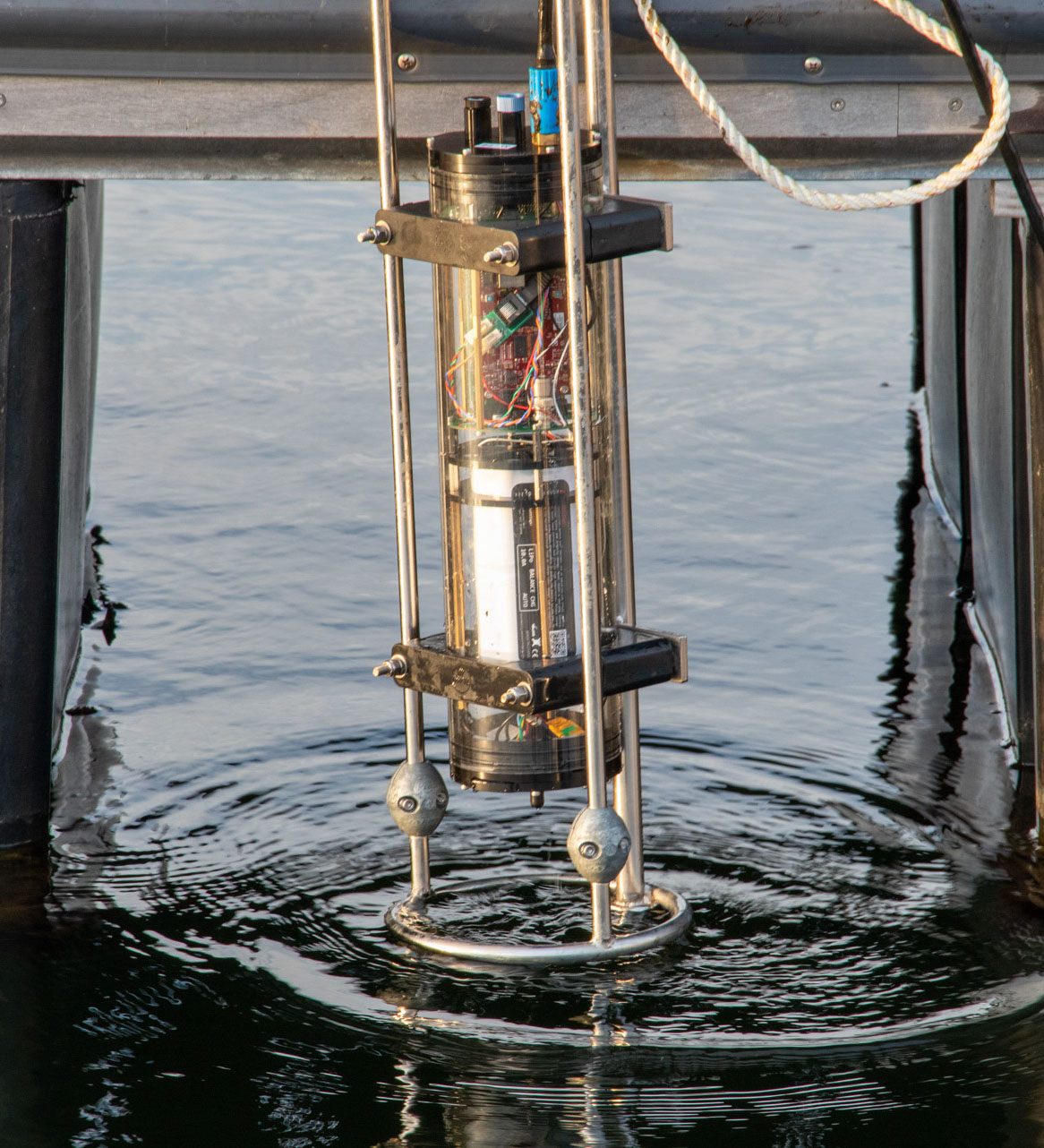}
     \end{subfigure}
     \caption{Deployed software-defined networking platforms.}
     \label{fig:modemdepl}
    \figsp
\end{figure}

\emph{Acoustic modems} have been shown to be viable solutions for long-range communications, as radio and optical technologies cannot be used by devices more than a few meters apart underwater~\cite{SendraLJP16}.
Even if \emph{acoustic modems} have been available for decades, current devices are beset by a host of limitations, including low data rates, cumbersome deployment procedures, and inflexible architectures.
These constraints hinder the advancement of emerging underwater technologies that need real-time high-resolution data transmission~\cite{singh07robotics,batchelor20geomorphology}. 
Common applications include transmitting imaging data through video streaming or side-scan sonar from autonomous underwater vehicles for monitoring and control purposes. 
These applications necessitate communication links with substantially higher data rates than those required for simpler sensing tasks. 
Currently, the data throughput capabilities of underwater networks are inadequate for high-quality data streaming, prompting investigations into potential solutions~\cite{ribas10oceans,campagnaro20jmse}.
Consequently, numerous IoUT applications are currently unfeasible with existing technology, motivating research and development of new devices designed to surmount these limitations.

\textbf{Contribution}. This work expands the possibilities of underwater communications and networking by designing, implementing, and testing an innovative software-defined underwater networking platform (Fig.~\ref{fig:modemdepl}) that offers the following unique characteristics: (i) It can reconfigure parameters at all layers of the protocol stack at run-time, to swiftly adapt to different settings and to changing environmental and communication conditions. This includes the physical layer---a typical bottleneck of current modems; (ii) it can facilitate high data rate applications (iii) it can be used standalone or integrated into other devices (e.g., underwater drones); (iv) it supports diverse set of tasks including dataset generation, channel estimation, Orthogonal Frequency-Division Multiplexing (OFDM) link establishment, bidirectional links using link layer protocols, and running third-party communication software, such as the JANUS standard for underwater communication. In addition, thanks to its unique hardware and software architecture, the proposed platform can be expanded to operate with different physical front ends, e.g., acoustic or optical as well as supporting MIMO capabilities via multiple transmitter and receiver chains.

To demonstrate the capabilities of our platform, we endow it with an acoustic front end and we deploy multiple prototypes at ocean for experiments (Fig.~\ref{fig:modemdepl}).
In particular, we show that we can use our devices to characterize the acoustic channel by determining the channel impulse response and frequency response.
We then show link establishment and investigate the quality of the link between two nodes exchanging OFDM packets. 
Our platform can be configured to achieve data rates that are up to $150$~kbit/s, which is more than twice the data rate of the fastest commercial underwater acoustic modem.
We finally show the ease of use of our platform with third-party software by installing and using the JANUS standard for underwater communication~\cite{PotterAGZNM14}.

The paper is structured as follows: Section~\ref{sec:soa} reviews existing platforms; Section~\ref{sec:design} details our design; Section~\ref{sec:prototyping} covers platform prototyping; Section~\ref{sec:exp_results} presents experimental results, and Section~\ref{sec:conclusions} concludes the paper.

\section{State-of-the-Art}
\label{sec:soa}

Despite the current absence of established standards and infrastructure, endeavors aimed at developing efficient underwater networks are expanding. 
A range of commercial options and research initiatives based on acoustic or visible light technology are presently underway. 
Commercial platforms generally prioritize the reliable transmission of data amongst devices made by the same manufacturer, whereas research platforms typically strive for open-source and flexible designs.
In this section, we provide an overview of the state-of-the-art on acoustic modems.
(Underwater platforms for visible light communications are still in their infancy and beyond the direct scope of this work. The interested reader is referred to~\cite{CaiMZ22-optical}).

\textbf{Commercial Acoustic Modems}.
In the realm of higher data rates, Evologics has developed high-speed, mid-range devices~\cite{evologics} that use a 60 kHz bandwidth and an omnidirectional transducer beam pattern. These devices reach a maximum range of 300 meters with data rates of up to $62.5$~kbit/s.
At lower data rates, Teledyne Marine's Benthos Acoustic Telemetry Modems offer a viable solution~\cite{teledyne}.
The Benthos have three working modalities: Low Frequency, reaching up to 6km, Medium Frequency, reaching up to 4km, and Band C, reaching up to 2 km.
They achieve data rates of $15.36$~kbit/s with a PSK transmit modulation.
The Benthos modems are compatible with omnidirectional and directional transducers and are designed to allow the usage of the JANUS standard.
Other modems with similar data rates include the Popoto Acoustic Modems~\cite{popoto}, which can achieve data rates of 10.24 kbit/s and also offer a JANUS-compliant version, and the Subnero~\cite{subnero} research modem, which employs PSK-OFDM and FH-BFSK modulations and supports JANUS.
The Subnero modem uses a $12$~kHz bandwidth and is able to achieve 15 kbit/s.
Comparable to the Subnero modem we also find the Sonardyne acoustic modems~\cite{sonardyne}, offering data rates up to $9$~kbit/s, and the Modem Embedable from Kongsberg~\cite{konsberg}, 
that uses a 9 kHz bandwidth and achieves data rates up to 6 kbit/s; it can mount several transducers.
Commercial modems still have two significant limitations: (i) their reliance on proprietary protocol stacks restricts the development of heterogeneous networks, and (ii) their performance is confined to predefined functionalities.

\textbf{Research Acoustic Modems}.
One promising research work is the acoustic modem presented by Mangione et al.~\cite{Mangione21}. The modem is developed based on the open-source Red Pitaya board and uses the liquid-dsp~\cite{liquid} library. 
Mangione's modem runs an OFDM adaptive system and uses a central frequency of 16 kHz; it delivers data rates up to $20$~kbit/s and is compatible with JANUS.
Another notable effort is currently being made by Campagnaro et al.~\cite{Campagnaro23}. They are developing a low-cost software-defined acoustic modem made only with off-the-shelf components. 
Campagnaro's modem uses a carrier frequency of 50 kHz and, for DSSS, a bandwidth of 30 kHz. When no Forward Error Correction is used, the raw bit rate at the physical layer is $1.5$~kbit/s with BPSK, and $240$~bit/s for DSSS.
Another relevant research effort is the ahoi modem~\cite{ahoi}. The ahoi modem is an open-source project and uses a $62.5$~kHz bandwidth; it offers data rates up to 4.7 kbit/s, demonstrated with a \SI{150}{\meter} link.
Finally, early versions of the SEANet networking platform presented in ~\cite{Demirors15WUWNet, Demirors16WUWNet2, Demirors18} offer promising reconfiguration and data rate capabilities including demonstrations of real-time physical layer reconfiguration.
While there have been multiple promising research acoustic modems, the advancement of an easily deployable, high data rate, plug-and-play, and task-diverse software-defined underwater networking platform remains in its infancy.

\section{Platform Design}
\label{sec:design}
In this section, we discuss the design of the three main components of the SEANet platform: The hardware, the software, and the mechanical parts that compose the enclosure. 
We then discuss some of the opportunities given by our modular design.

\subsection{Hardware Design}
\begin{figure}
    \centering
    \includegraphics[width=0.7\linewidth]{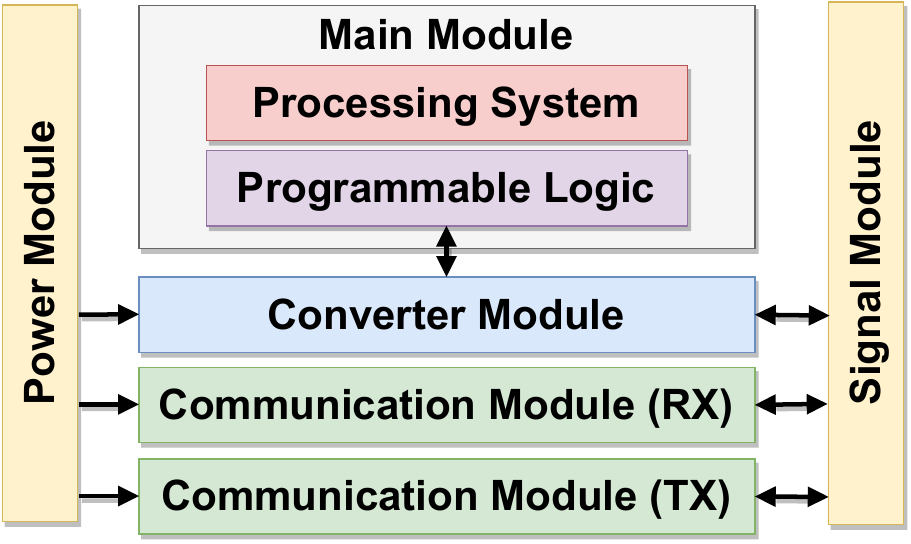}
    \caption{Hardware architecture of the SEANet networking platform.}
    \label{fig:system_architecture}
    \figsp
\end{figure}

The hardware architecture of the SEANet networking platform is shown in Fig. \ref{fig:system_architecture}.
It consists of four core modules: the main module, the converter module, and two communication modules.
Two auxiliary modules for power and signal interconnects complete the design.

There are three main challenges that the hardware design needs to address: (i) the electrical needs of the transducer need to be met, (ii) it needs to be able to operate at ultrasonic frequencies, and (iii) it needs to fit into an enclosure suitable for underwater deployment and operations. 

\subsubsection{Main Module}
This is the computational core of the SEANet networking platform. It performs signal processing, runs the software protocol stack, and the real-time adaptation and reconfiguration. It is based on a System-on-Chip (SoC) architecture that consists of an ARM-based processor and a field programmable gate array (FPGA). 
The Processing System (PS) is responsible for all the upper layers of the protocol stack, while the Programmable Logic (PL) is used to implement computationally intensive physical layer operations in real-time.
\subsubsection{Converter Module}

The design of the converter module is motivated by the requirement to support diverse tasks in the underwater communications platform and to seamlessly integrate with various frontends. 
This necessitates the use of dedicated high-performance wideband data converters, which were previously unavailable as a main module option and are now incorporated within a specialized converter module.
The converter module interfaces the main module with the rest of the modules. 
It has analog-to-digital (ADC) and digital-to-analog (DAC) data converters that provide analog interfaces to and from communication modules. In addition, it has voltage regulators for the main module, clock sources, and digital I/O ports. Since the platform is based on a direct conversion architecture and relies on digital mixers, to support wide range of frequencies used for underwater communications, fast data converters are chosen. 
\subsubsection{Communication Modules}

The analog frontends of the platform must be tailored to suit specific applications, similar to commercial SDR platforms used in terrestrial networks. 
They should be engineered to meet performance criteria such as output power, bandwidth, and receiver sensitivity for the particular transducers they will utilize. 
This adaptability allows them to be configured for various tasks, including localization and acoustic sensing, and to function unidirectionally as either a beacon or a sensor. 
In this design, our focus is on wideband, high-frequency underwater acoustic communications applications.

\begin{figure}[h]
    \centering
    \includegraphics[width=.9\linewidth]{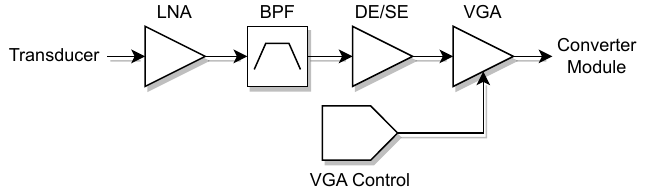}
    \caption{Communication module preamplifier block diagram.}
    \label{fig:preamp_bd}
\end{figure}

These frontends perform amplification and filtering of incoming and outgoing signals. In our system, we consider one receiver module (RX) housing the preamplifier, and one transmission module (TX) housing a power amplifier. 
The preamplifier consists of three stages: a low noise amplifier (LNA), a filter, and a variable gain amplifier (VGA) as shown in Fig.~\ref{fig:preamp_bd}. 
The primary function of the first stage is impedance matching to the transducer. The second stage is an optional band-pass filter that removes out-of-band noise. The last stage consists of a differential driver and a VGA that provides further amplification and produces a differential output for the converter module ADC inputs. 

\begin{figure}[h]
    \centering
    \includegraphics[width=0.82\linewidth]{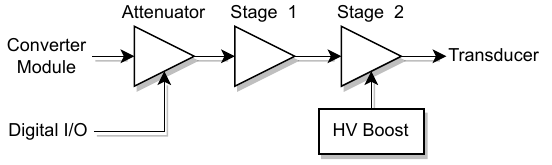}
    \caption{Communication module power amplifier block diagram.}
    \label{fig:pa_bd}
\end{figure}

The power amplifier consists of a digitally programmable attenuator followed by a two-stage amplifier block as illustrated in Fig.~\ref{fig:pa_bd}. The module is designed to drive transducers, which are capacitive loads requiring high voltage stimulus. The amplification stages are based on integrated circuit amplifiers with large gain and capacitive drive capability. The programmable attenuator allows transmitting levels to be adjusted.

In this design, a high voltage power amplifier is preferred instead of a high current output amplifier followed by a matching transformer. The latter option requires a custom transformer for the specific transducer at a given frequency range that limits compatibility and adds cost and complexity. 

\subsubsection{Power Module}

The platform design features a centralized power module, which supplies ready-to-use power rails to other modules, thereby streamlining the overall design process. 
This strategy not only optimizes the use of available board space for additional modules but also effectively separates noise-emitting components from critical signal pathways, enhancing the system's overall performance.
The power module is an interface to the battery unit and powers each module with different voltages using onboard voltage regulators. %
The module includes filters to reduce power supply noise and protection circuitry that continuously monitors the voltage and current levels delivered to the platform.
\subsubsection{Signal Module}

A separate signal module is included to provide expandability and support multi-channel operation.
To provide this level of flexibility, a large number of signals need to be routed across different modules to create one-to-many connections.
Using connectors on modules with cable assemblies is difficult to design and manage at this scale, so we use the following approach instead.
The signal module routes digital and analog signals between the converter and communication modules and the transducer.
As in the power module, each signal is accessible from any board in the stack. 
The signal module mounts reconfigurable board edge connectors to provide flexibility, e.g., it is possible to have more than one converter module attached to it. 
The signal module also has a connector for the transducer interface. This connection is driven by a duplexer circuit whose inputs can be switched between the power amplifier and preamplifier channels to allow a single transducer to be used for bidirectional communication. 
Even though transmit and receive chains can operate simultaneously, a half-duplex operation is implemented mainly due to the high cost of high-bandwidth acoustic transducers.
\begin{figure}[h]
    \centering
    \includegraphics[width=0.9\linewidth]{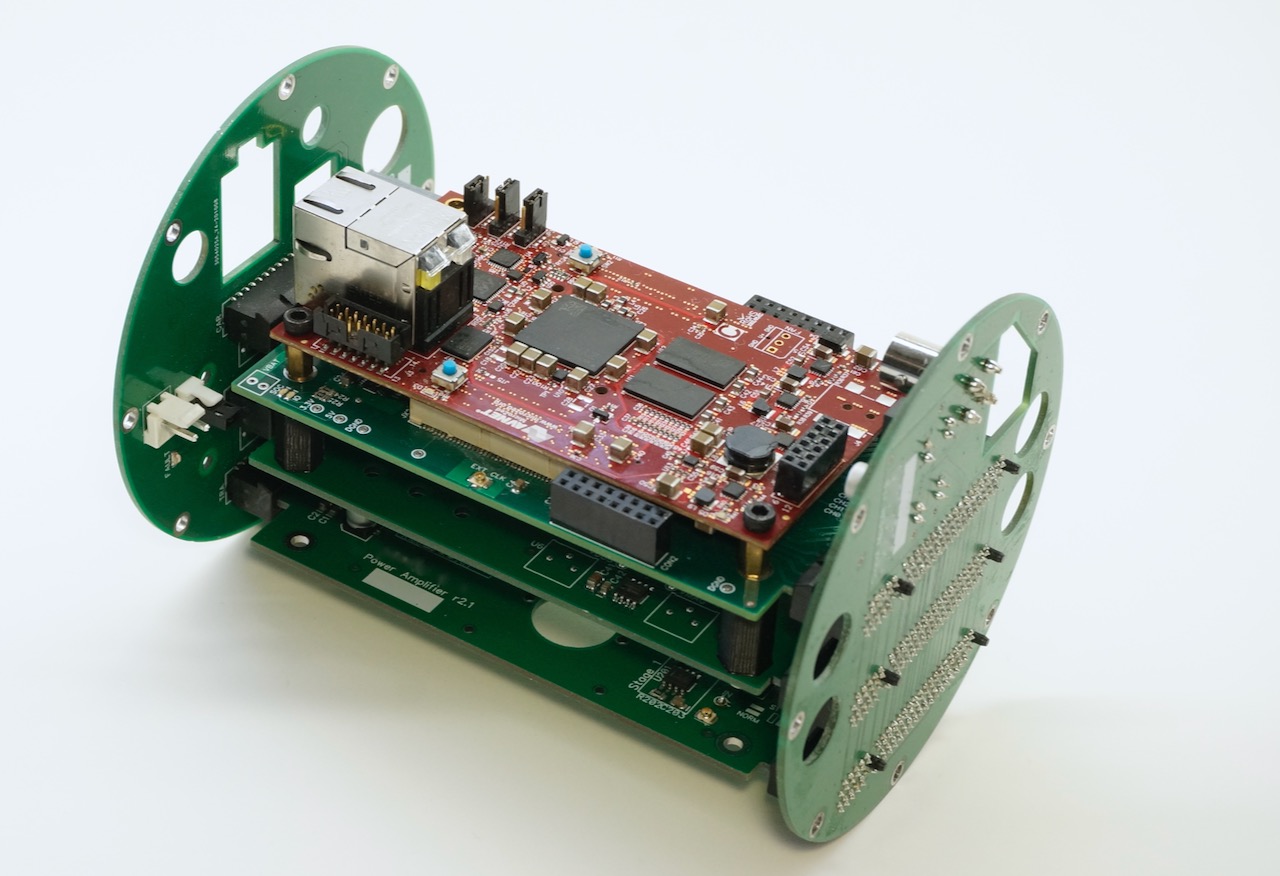}
    \caption{Assembled hardware of the SEANet Platform.}
    \label{fig:hardware}
    \figsp
\end{figure}

\subsection{Software Design}
\label{sec:software}
The development of the software is split between the processing system (PS) and the programmable logic (PL).
The signal-processing computations of the physical layer are developed in the FPGA while the processor runs the rest of the protocol stack in parallel.
Despite the relatively low achievable rates underwater, the utilization of a direct conversion architecture requires DSP blocks to handle continuous data at rates in the order of 10 Mbit/s. In particular, the implementation of digital filters with numerous taps and large Fast Fourier Transforms (FFT) may face limitations when confined to the PS due to latency constraints. Offloading these tasks to the PL leads to a noteworthy decrease in design complexity.

\subsubsection{Programmable Logic Design}

The PL is designed to offer a reconfigurable physical layer that is able to interact with the higher layers of the protocol stack, implemented in the PS. 
The design comprises two main functionalities: waveform streaming and recording, and zero-padded OFDM (ZP-OFDM) transceiving. 
The former implements generic software-defined radio functionalities and relies on the processing system for signal processing. 
ZP-OFDM is an integrated physical layer implementation used to transmit and receive packets of binary data over the acoustic interface. 
The waveform recorder is used as a basis for implementing physical layer designs as it includes common processing system interfaces for data streaming and hardware interface blocks.
\subsubsection{Processing System Design}
The PS is responsible for providing a user interface, implementing higher layers of the protocol stack (e.g., MAC, Network, Transport, and Application), and transferring data and configuration packets to/from the PL using a DMA Driver Wrapper. The PS also enables the reconfiguration of parameters of the physical layer like bandwidth and center frequency or switching to a different physical layer altogether. The PS must also be computationally light since it is running on an embedded device.

\textbf{User Interface.} 
The user interface (UI) offers the operator of the system the possibility to perform several tasks, e.g., send a file, or read from a device. It also allows the user to input a variety of parameters, e.g., amplifier gain levels.
The UI also allows to switch between the protocols available at each layer of the protocol stack. 

\textbf{Protocol Stack.} The protocol stack is designed to incorporate, from top to bottom, the Application Layer, the Transport Layer, the Network Layer, and the Link Layer. 
As previously specified, the Physical Layer is implemented in the PL.
Each layer is an independent module and each module can be developed with a different programming language. 
The layers are connected with each other using synchronous and asynchronous pipes.
The modular structure allows for easy expansion, modification, or replacement of each module.

\textbf{DMA Driver Wrapper.} The wrapper for the Xilinx AXI DMA Driver allows for usage of the Xilinx kernel driver from the user space.
The driver handles transfers of data between the PS and the PL through the DMA. The DMA allows for data transfers between peripherals and the main memory without passing through the system processor.
The efficient implementation of the driver allows to read/write in kernel-space memory with user-space functions.

\textbf{Runtime PL Reconfiguration.} 
It is possible to change the parameters of the physical layer by modifying registers defined in the memory map of the PS. 
It is also possible to change the running physical layer using the FPGA manager for flashing a different PL bitstream within the OS without a reboot.

\subsection{Mechanical Design}
The following design principles drive the realization of the SEANet networking platform's enclosure, which is shown in Fig.~\ref{fig:modem_mech}: (i) to prevent electrical failures, the enclosure is watertight and resistant to water pressures at depths of at least 100~m, (ii) it is also lightweight to achieve a multi-node deployment without machine aid, (iii) since the platforms need to be usable for several deployments their outside must be resilient to the sea conditions, but also (iv) they need to have the possibility to be disassembled and reassembled to allow researchers to try different solutions in terms of electronics. To allow underwater communication, there must be water-tight openings (v) for transducers, (vi) for a data cable to allow for a connection with the user, and (vii) for a mechanical switch that can be used in an extreme situation to restart the system without opening the enclosure. Finally, the platforms (viii) need attachment places to be fixed to the seafloor or at a certain depth. 
\subsection{Modular Design}
\label{sec:modular_design}
The SEANet networking platform is flexibly designed to use different frontends. The entire hardware design is based on swappable hardware modules connected by standard interfaces. 

\textbf{Frequency Band Adaptation.}
As a general-purpose software-defined radio platform for underwater networks with large bandwidths, the SEANet networking platform can be utilized in many applications with minor modifications. 
For example, if lower frequencies are preferred to communicate over long distances, due to frequency-dependent propagation loss~\cite{melodia2013advances}, custom communication modules and specific transducers can be integrated into the platform to meet performance specifications.

\textbf{MIMO.}
The proposed platform architecture can be extended to use the multiple-input and multiple-output (MIMO) technique by adding multiple transmitter and receiver chains. The resulting system can be used to implement spatial multiplexing or transmit diversity to improve the throughput or reliability of the acoustic link. With multiple analog paths on both converter and communication modules, the signal module can be used as a crossbar to route input and output signals to multiple communication modules. With slight modifications to the signal and power modules to increase available connections, the current design of the SEANet platform can support~4x4 MIMO.

\textbf{Visible Light Communications.}
For high data rate underwater links at shorter distances, visible light communication (VLC)--based modems can be used. 
Typical underwater VLC systems utilize intensity modulation and direct detection, operate over larger bandwidths than acoustic modems, and require different amplifier topologies for photodetectors and LEDs/lasers~\cite{7593257}.

The software-based signal processing architecture of the SEANet networking platform can be used for implementing VLC physical layers. At the same time, our platform accepts custom analog VLC frontends. 
Detailed information about this adaptation including VLC-specific hardware and communication schemes can be found in~\cite{enhos21secon}. The proposed system demonstrates bit error rates less than $1\times10^{-6}$ and data rates $1$~Mbit/s in experiments at sea.

\section{Prototyping} 
\label{sec:prototyping}

This section outlines the implementation of the design introduced in Section~\ref{sec:design}.
We begin with the hardware implementation, then move on to the software components.
Lastly, we address the development of the mechanical structure.

\subsection{Hardware Implementation}
\subsubsection{Main Module}

The main module executes the software architecture design while maintaining energy efficiency to operate on battery power.
It should include all necessary interfaces for connectivity and sensors, although ports for user peripherals like displays or inputs are optional. 
Also, the module must be compact to fit within the size limitations of our underwater enclosure. 
We favor commercial computation modules due to their cost-effectiveness and the extensive community support available for both software and hardware development.
The main module is realized using a MicroZed system-on-module (SoM) board. The MicroZed is a low-cost development board based on the Xilinx Zynq-7000 All Programmable SoC that integrates an ARM-based processor with a programmable FPGA. The SoM has two I/O headers that provide connection to two I/O banks on the programmable logic (PL). When the main module is plugged into a carrier card, the I/O headers are accessible in a way that is dependent on the carrier design.
\subsubsection{Converter Module}

The core component of this module is the high-speed data converters that implement the direct conversion architecture.
To reduce distortion from aliasing, these converters require high sampling rates exceeding the Nyquist rate of the ultrasonic frequencies utilized in acoustic communications.
In addition to the large bandwidth requirement, they need to be high-precision to operate with low noise levels and have a good dynamic range to be able to receive transmissions from both nearby and far away transmitters.
Most common data converter products for audio (high-precision, low-speed) or communication (high-speed, low-precision) applications do not meet these specific requirements.
Therefore we have implemented this module as follows.

The data converter section of the converter module is based on LTC1740 6~Msps 14-bit ADC and LTC1668 50~Msps 16-bit DAC. A temperature-compensated crystal oscillator is present on the module to be used as the clock source for the ADC, DAC, and DSP blocks in the PL design. 
The power supply for the main module is implemented with high-efficiency switching regulators based on LTC3624.
The design of the power supply for this module is challenging as most of the voltage regulators have switching frequencies that overlap with the spectrum commonly used for underwater acoustic communications. 
The resulting harmonic content of the switching noise causes poor performance, especially in multi-carrier communications. For these reasons, this board is implemented with a four-layer printed circuit board, analog and digital sections with isolated ground planes to keep high-speed digital traces away from analog signals, and to have separate power rails sourced from the power module for ADC and DAC.
\subsubsection{Communication Modules}

The main difficulty in deploying the communication modules lies in interfacing with transducers. 
Standard amplifier configurations typically used in communications frequently face difficulties with these capacitive devices, leading to poor SNR performance. 
Although there are commercially available amplifiers designed for transducers operating in the ultrasonic frequency range, many are too large for our enclosure or require power inputs that do not match our platform specifications. 
Therefore, to achieve better integration, custom amplifier chains are utilized for these modules.

The preamplifier module consists of three stages: a low noise amplifier (LNA), a filter, and a variable gain amplifier (VGA). The first stage is an integrated impedance matching and amplification stage with \SI{40}{\decibel} gain implemented with a low noise op-amp AD8067 in non-inverting configuration.
The input impedance of the amplifier is high enough to provide efficient voltage transfer from the transducer over the usable frequency range. This stage determines the signal-to-noise performance of the receiver section as the noise level at the output of this section is further amplified by the following stages.
The signal level at the output of the LNA may be too small for packet detection for longer link distances due to transmission loss. An AD8330 VGA follows the LNA to provide adjustable amplification up to \SI{50}{\decibel}.
The gain interface of the VGA, adjustable from the PS, is analog and it is driven by MCP4812 10-bit SPI DAC. 

The preamplifier module also has a 4-stage active bandpass filter with cut-off frequencies at \SI{10}{\kilo\hertz} and \SI{250}{\kilo\hertz}. The filter is designed to block out-of-band acoustic and electrical noise sources while allowing a large spectrum. The filter can be bypassed with a jumper for low-frequency operation. 

The power amplifier module is based on an integrated operational amplifier PA340 and a voltage boost circuit. The gain-bandwidth product of the PA340 is not large enough to amplify the DAC signal levels at \SI{200}{\kilo\hertz} therefore a preceding amplification stage based on OPA192 operational amplifier is implemented to distribute the total gain. The output level is set through MAX5420 which is used as an attenuator in the first stage. The bipolar \SI{\pm70}{\volt} supply rails is generated with an LT3511 implementing isolated flyback converter topology.

\subsubsection{Power Module}

The modules described above need multiple voltage rails that differ in current requirements and noise specifications. 
Commercial power supply modules, while meeting the necessary voltage and current specifications, are unsuitable for two primary reasons. 
First, their noise characteristics at ultrasonic frequencies generate in-band noise, which impairs the SNR at the receiver. 
Second, they fail to conform to the mechanical specifications once assembled. 
Consequently, a custom power module has been developed that functions both as a structural element and a power source, deriving power rails directly from the battery supply.

The module utilizes low-noise linear voltage regulators to generate voltage rails for the data conversion modules and the preamplifier.
The system is protected against overvoltage, undervoltage, and short-circuit faults to maintain safe operation with high-capacity lithium-ion batteries. 
An external power switch can turn off or reboot the system without opening the enclosure. Both the power module and the signal module utilize card edge connectors used for the PCI Express bus standard which are ubiquitous, low-cost, and robust. The modules however have custom pinouts and form factors.

\subsubsection{Signal Module}

The signal module requires densely packed connectors for all analog and digital signal connections across the boards, while also accommodating future expansions with additional communication modules. 
The connectors must be durable and have a pitch that eases the hardware design of other modules. 
Although common coaxial connectors like SMA are sturdy, they necessitate substantial clearance for mating, which complicates the mechanical design. 
Smaller coaxial connectors, while less bulky, tend to detach easily and require labor-intensive inspection and maintenance. 
A more suitable approach involves adopting a unified single connector design that doubles as a mounting component for other modules.
This design enables the signal module to also function as a structural element, securing the other modules in place.
Our signal module has 98-pin board edge connectors similar to the power module.
For digital I/O, 20 high-speed differential pin pairs are routed to the main module and separated from each other with ground connections. 
The analog I/O signals have four types of connections: preamplifier to ADC, DAC to power amplifier, transducer to preamplifier, and transducer to power amplifier. Each type of connection is replicated four times, forming four individual channels, and that at modules through jumpers. 

The transducer input and output signals are connected to a duplexer circuit which is an SPDT switch. This switch element needs to withstand high voltages produced by the power amplifier and have a low loss to preserve the received voltage from the transducer. Adequate isolation to protect the preamplifier input is provided with a reed relay, which is preferred over solid-state switches. The resulting switching latency is in the order of a few milliseconds which usually does not exceed propagation delay for typical link distances. 

The assembled modules form a cylindrical structure that is \SI{135}{\milli\meter} long and \SI{100}{\milli\meter} wide (Fig.~\ref{fig:hardware}). 

\subsection{Software Implementation}
The software implementation is divided into programmable logic implementation and processing system implementation.
The core challenge we faced was the lack of plug-and-play solutions able to address
the fundamental requirements of our system, including, a reconfigurable and modular protocol stack for underwater networks able to continuously stream data between the physical layer (in PL), and the higher layers (in PS) of the protocol stack while offering an intuitive user interface.

\subsubsection{PL Implementation}
\label{sec:pl_implementation}
In this section, we discuss the implementation of the two basic PL blocks: the waveform streamer and recorder, and the ZP-OFDM transceiver.

\begin{figure}
    \centering
    \includegraphics[width=\linewidth]{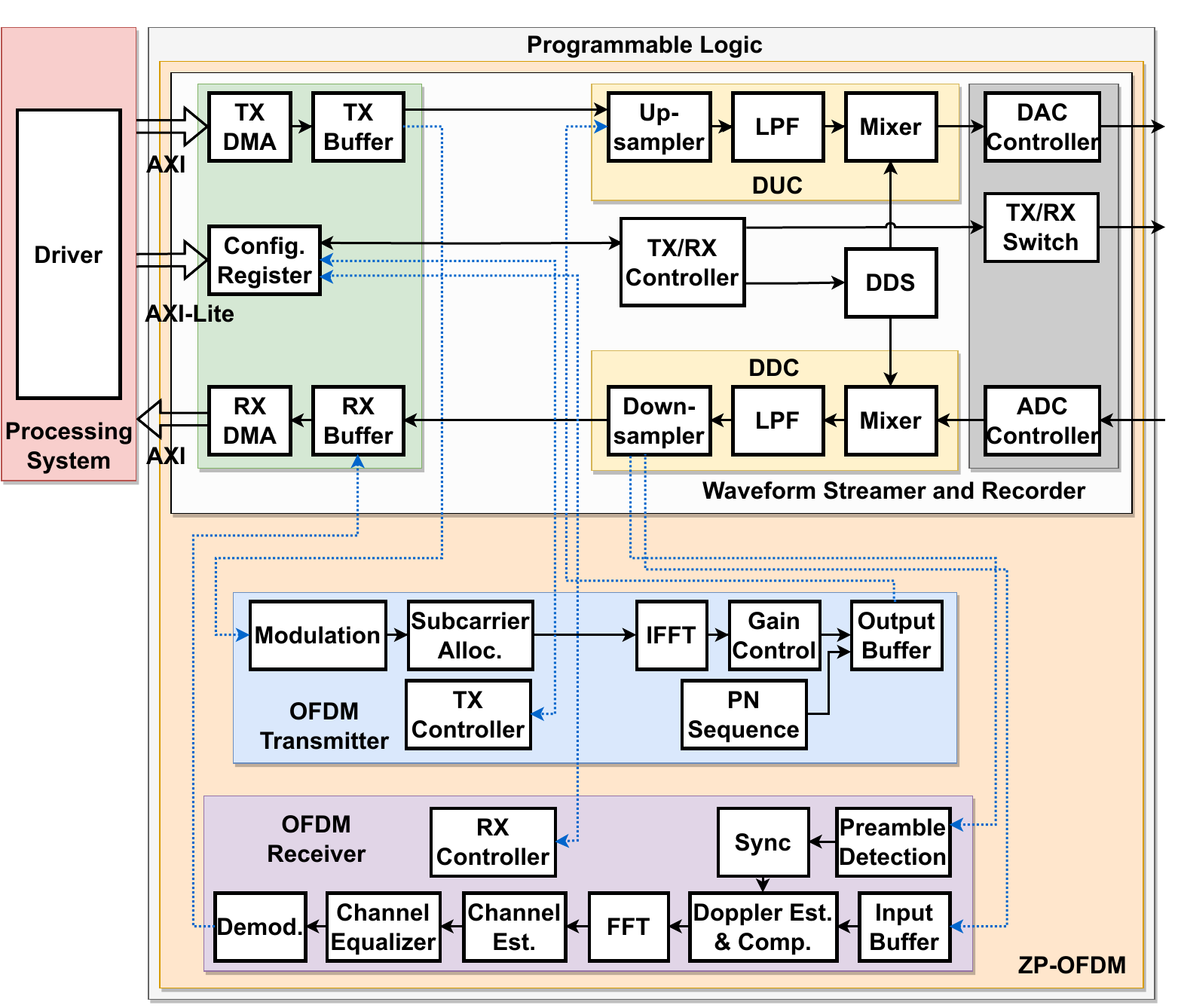}
    \caption{The top part of the figure identifies the waveform streamer and recorder block diagram. The ZP-OFDM design (lower part) is obtained by adding the OFDM transmitter and receiver to the waveform streamer/recorder.}
    \label{fig:pl_zpofdm_bd}
    \figsp
\end{figure}

\textbf{Waveform Streamer and Recorder.}
To enable physical layer development, a waveform streamer and recorder design is implemented in the PL which allows the device driver to transmit/receive baseband samples directly between PL and user space. This mode, along with the accompanying software, provides functionalities typically offered by many commercial software-defined radio platforms. The block diagram for the design is given in the upper section of Fig.~\ref{fig:pl_zpofdm_bd}.

On the PS interface, the design incorporates direct memory access (DMA) controllers and configuration registers that use AXI and AXI-Lite interfaces respectively. The DMA controllers are buffered for rate matching between PL and PS. The design also features a real-time configurable digital down converter (DDC) and digital up converter (DUC) which can be enabled separately to be used with I/Q baseband samples. In this operation mode, the center frequency is set using a direct digital synthesis (DDS) block. On the TX chain, the upconverted signal from the DUC or the passband signal from the TX buffer is routed to the DAC controller for transmission. On the RX chain, samples are received from the ADC controller and either written to the RX buffer or downconverted with the DDC. These controllers are responsible for generating control signals for ADC and DAC hardware and configuring sample rates.
Finally, hardware control signals for duplexer and communication modules are generated by controller blocks. 

\textbf{ZP-OFDM Transceiver.}
The second PL design includes a zero-padded (ZP)-OFDM transceiver. A block-level representation of it is depicted in Fig.~\ref{fig:pl_zpofdm_bd}. 
On the transmitter chain, information bits are received from PS through an AXI interface with TX DMA into the PL. The inputted information bits are mapped into symbols according to the selected modulation scheme. The generated data symbols are then allocated to subcarriers alongside pilot and null symbols to form the OFDM symbols. The formed OFDM symbols are then forwarded to the IFFT block to be translated to the time-domain. The gain of time domain symbols is adjusted with a gain control unit and formed into a packet format, which includes a preamble (PN sequence) and a selected number of OFDM symbols. The generated packets are later up-converted to the passband frequency by a digital up-converter (DUC) module and sent to the DAC unit through the DAC controller block. 

On the receiver chain, the received signals from the ADC unit are fed into a digital down-converter (DDC) module through the ADC controller block. The received signals are filtered to eliminate DC offset and out-of-band noise and down-converted to the baseband by the DDC module. The baseband signals are then processed by a preamble detection block which performs autocorrelation-based packet detection and time synchronization.
Each packet is then partitioned into OFDM symbols. Then symbols are translated into the frequency domain with FFT, and passed through the blocks performing Doppler scale estimation and compensation, pilot-tone based channel estimation, zero-forcing (ZF) channel equalization, and symbol detection at the OFDM receiver module. The detected symbols are later mapped into bits based on the selected modulation scheme and fed to the PS through an AXI interface with RX DMA. In addition to the blocks that are realizing a ZP-OFDM scheme transceiver logic, the PL design also includes configuration registers that are responsible for storing physical layer configuration parameters set by the PS through an AXI-Lite interface. The stored parameters are later used by controller blocks (i.e., TX/RX, TX, RX), which are also responsible for coordinating physical layer operations implemented by the other blocks of the chain through finite state machines (FSMs), to reconfigure the physical layer configurations. The current prototype allows multiple parameters to be reconfigured in real-time, e.g., modulation, guard time, subcarrier mapping, number of subcarriers, bandwidth, and carrier frequency.

\subsubsection{PS Implementation}
The processing system development is based on a custom Linux distribution for embedded systems. The development environment is built upon the Yocto Project and compiled with OpenEmbedded~\cite{yocto}.
Among all the other functionalities, the Linux distribution supports standard interfaces (i.e., Ethernet, USB, UART, CAN, EBI/EMI, I²C, MMC/SD/SDIO, and SPI) which are available on the platform.
As mentioned in section~\ref{sec:design}, the PS is where the protocol stack is implemented. 

The \textbf{Application Layer} allows us to perform several tasks. When performing data communication, the application layer can show online statistics like bit error rate and signal-to-noise ratio. 
The \textbf{Network Layer} and the \textbf{Transport Layer} are currently transparent.
The \textbf{MAC Layer} is implemented as a FSM that prioritizes incoming packets over the packets that need to be transmitted. The MAC Layer's header includes source and destination IDs and a CRC to check the correctness of the received frame. 
The layers of the protocol stack in the PS exchange data using named pipes (FIFO special file) in blocking mode to synchronize the layers over them. 
For data transfers from PL to PS, the MAC layer uses Xilinx's AXI DMA driver in synchronous mode.
The transfer from PS to PL is also blocking, thus the process is blocked on the write operation until the data has been correctly received in the PL.

\subsection{Mechanical Implementation}

\begin{figure}
    \centering
    \includegraphics[width=0.9\linewidth]{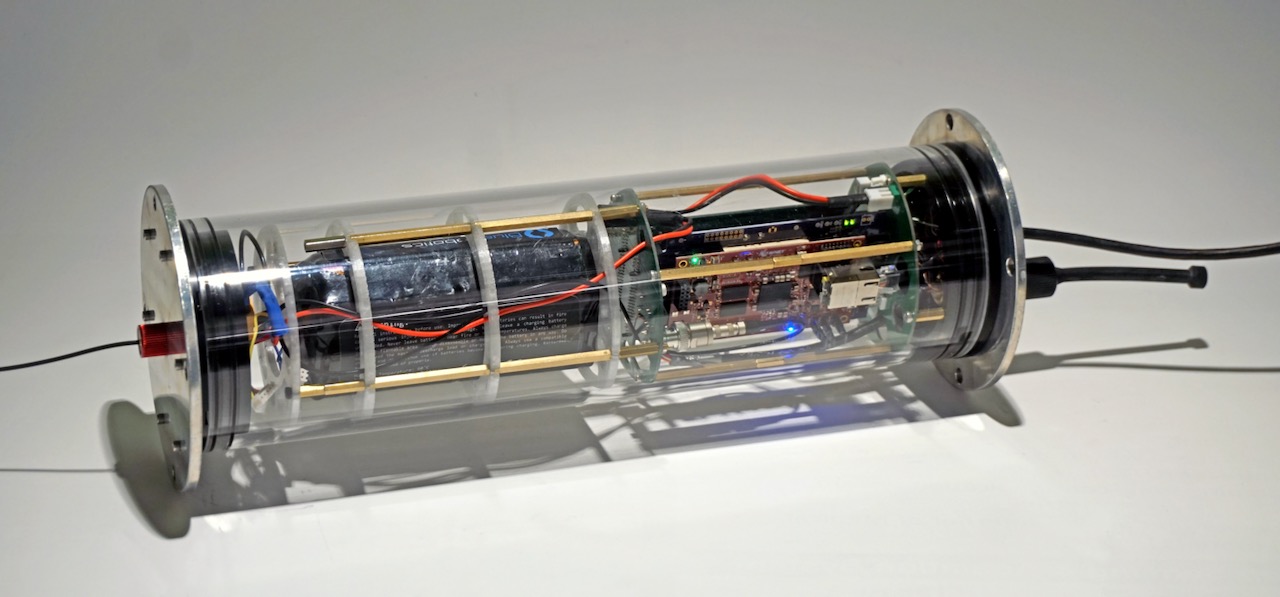}
    \caption{Fully deployable SEANet networking platform. The battery pack is included in the enclosure (on the left).}
    \label{fig:modem_mech}    
\end{figure}

For the mechanical implementation, our priority was to select an enclosure that met our design requirements.
As enclosure size increases, its pressure rating typically decreases, which requires more robust construction to achieve the necessary pressure tolerance. 
While off-the-shelf enclosures made from materials like aluminum or titanium are available, they tend to be heavier, offer higher depth ratings, and are considerably more expensive.
Although these enclosures might be suitable for specific applications, they add complexity, requiring specialized maintenance tools and additional buoyancy to offset their weight.
Therefore, we decided on a lighter alternative for our enclosure to fulfill our design goals.
The mechanical structure of the SEANet networking platform is shown in Fig.~\ref{fig:modem_mech}.

To realize the casing of the SEANet networking platform, we used a $4$~in diameter acrylic tube.
The tube is enclosed by two custom aluminum caps housing the transducer, ventilation port, power switch, and Ethernet connector.

The mechanical design of the SEANet platform allows for \textbf{standalone and integrated deployments}. In a standalone deployment, the platform can be: (i) moored to the bottom, (ii) attached to a smart buoy for long-range access~\cite{falleni2020design} or (iii) directly connected to a user through the Ethernet cable. As previously demonstrated, the light mechanical design of the platform allows it to be easily integrated into most commercial underwater vehicles~\cite{Unal22}. 

\section{Experimental Results}
\label{sec:exp_results}

In this section, we present an experimental evaluation of the proposed platform. Initially, we present results on characterization and verification of the platform's functionality. Subsequently, we showcase potential applications across different domains, highlighting both its versatility and potential.

\subsection{Platform Verification}

\begin{figure}[h]
	\includegraphics[width=\linewidth]{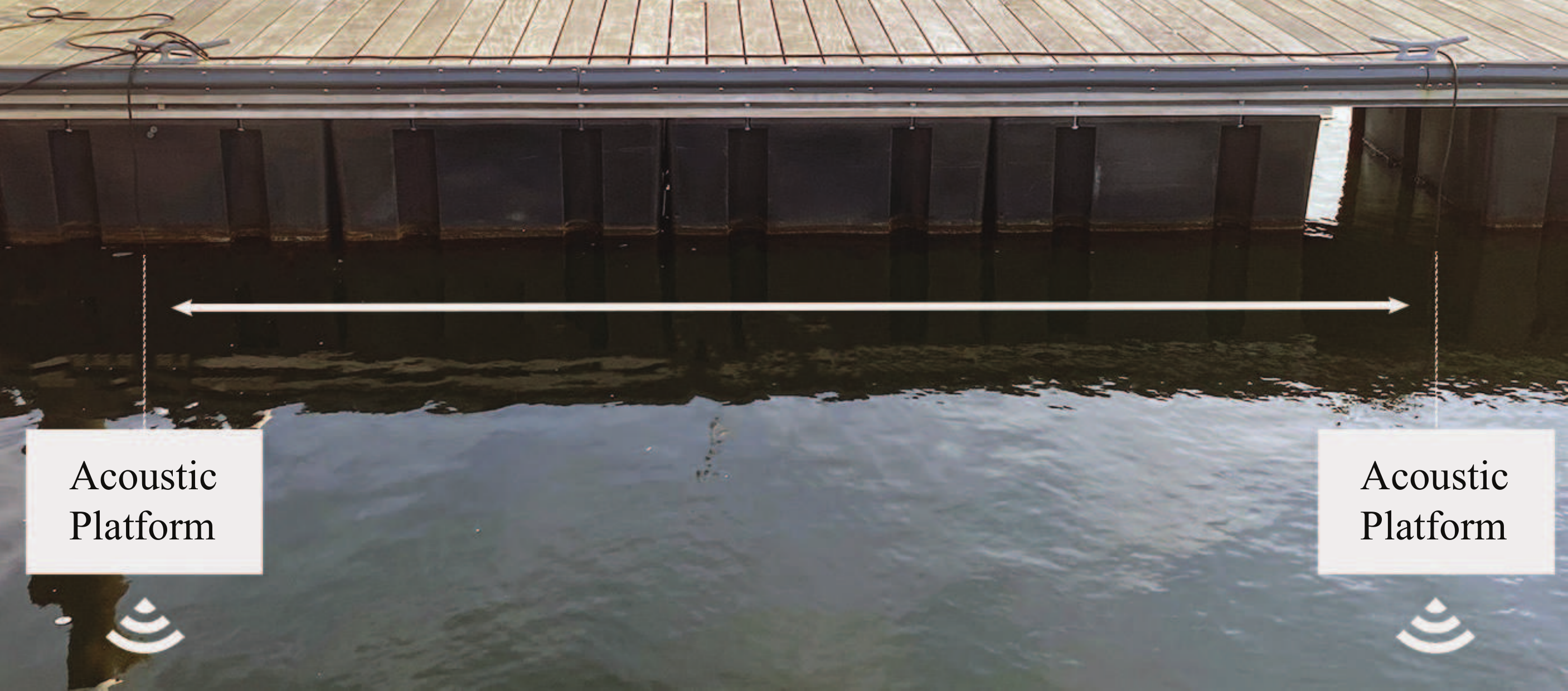}
	\caption{Marina deployment setup.}
	\label{fig:deploymentsc}
\end{figure}

Initially, we assess power consumption under different operational settings, followed by data streaming and recording tests to validate both capabilities for generating datasets as well as validating hardware and software operations.
We then establish an OFDM-based communication link, analyzing its performance across diverse modulations and bandwidths at different distances.
All tests are conducted in a marina setting (with a total depth of approximately 12 meters) using two prototype platforms deployed 5 meters apart at a depth of 3 meters (Fig.~\ref{fig:deploymentsc}).

\subsubsection{Power Consumption}\label{sec:power}
We evaluate the platform's power consumption in different operational modes, directly by measuring from the battery terminals. %
The platform consumes 5.97~W when idle and 6.11~W in receive mode.
When transmitting, the platform consume from a minimum of 6.4~W to a maximum of 9.2~W.
Considering the power consumed by the operations and the $18$~Ah capacity 4S lithium-ion battery mounted on each prototype, the expected life range goes from 44 hours when idle to 29 hours when in continuous transmission. 

\subsubsection{Data Collection and Channel Characterization}
\label{sec:exp_channel_response}

In this part of the experimental study, we focus on showcasing the proposed platform's ability to perform data collection both for generating datasets for data-driven AI/ML algorithms and characterizing the acoustic channel. 

\begin{figure}[h]
     \centering
     \begin{subfigure}[b]{\columnwidth}
         \centering
         \includegraphics[width=\linewidth]{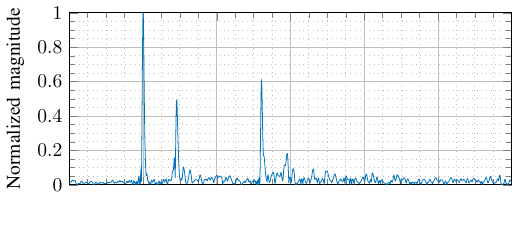}
         \vspace{-0.9cm}
     \end{subfigure}     
     \begin{subfigure}[b]{\columnwidth}
         \centering
         \includegraphics[width=\linewidth]{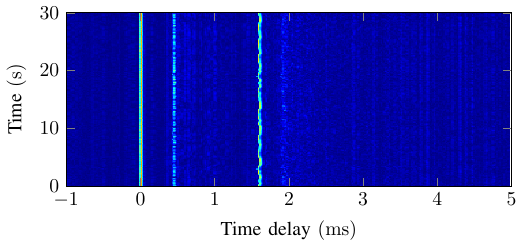}
     \end{subfigure}
     \caption{Channel Impulse Response: Normalized magnitude (top), and over time (bottom).}
     \label{fig:imp_resp}
\end{figure}

First, we generate a dataset from the acoustic channel by transmitting PN pulses at a \SI{100}{\kilo\hertz} center frequency over a \SI{100}{\kilo\hertz} bandwidth, continuously. Later on, we use this dataset to obtain the channel impulse response by autocorrelating PN pulses as described in \cite{vanWalree2013JOE}. Fig.~\ref{fig:imp_resp} (top) shows the response that includes a direct path and three reflection paths. Moreover, using the collected dataset, we obtain the variations of the channel response over a duration of 30~s, as shown in Fig.~\ref{fig:imp_resp} (bottom). Such datasets are important as they can be used both for designing conventional mode-based communication and networking protocols as well as novel data-driven approaches~\cite{otnes15joe,walree17joe}.

\begin{figure}[h]
    \centering
    \includegraphics[width=\linewidth]{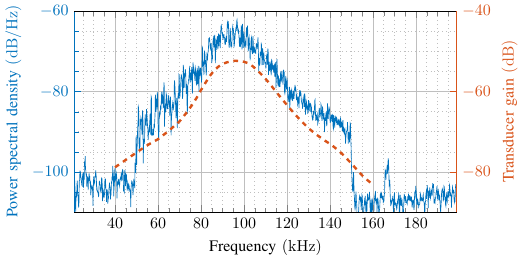}
    \caption{Measured frequency response (left axis) and combined gain of transmitter and receiver transducers (right axis).}
    \label{fig:freq_response}
\end{figure}

Finally, in this set of experiments, we measure the end-to-end frequency response of the system that is the combined performance of communication modules, transducers, and acoustic channel. We use linear frequency modulation (LFM) pulses that cover the frequency band to measure received power level. The received average power per frequency in the \SI{50}{\kilo\hertz}--\SI{150}{\kilo\hertz} band is illustrated in Fig.~\ref{fig:freq_response}.
In addition, the combined transducer response is overlaid on the plot with a red dashed line. The response is calculated using transmitting voltage response (TVR) and open circuit receiving response (OCRR) measurements extracted from the transducer calibration datasheets. The TVR and OCRR measurements correspond to the conversion between signal power and acoustic pressure for a given frequency, which are then combined to calculate gain values.

We observe that the received power spectral density in this band is mainly determined by the characteristics of the transducers. The peak around $166$~kHz is present even without transmission and therefore is considered noise and disregarded for this analysis. Note that the availability of wide bandwidth transducer options is limited, and they often come with significantly higher costs. Consequently, we opt for these transducers to demonstrate high-bandwidth use cases.

\subsubsection{Physical Layer}

In this part of the experimental study, we focus on the physical layer capabilities of the proposed platform. To that end, we first implement and demonstrate a ZP-OFDM communication link where we evaluate its Bit Error Rate (BER) performance at different Signal-to-Noise Ratio (SNR) levels. Consequently, to showcase the software-defined supported reconfiguration and optimization capabilities of the proposed platform, we focus on demonstrating high data rate communication links.

\begin{figure}[h]
    \centering
    \includegraphics[width=\linewidth]{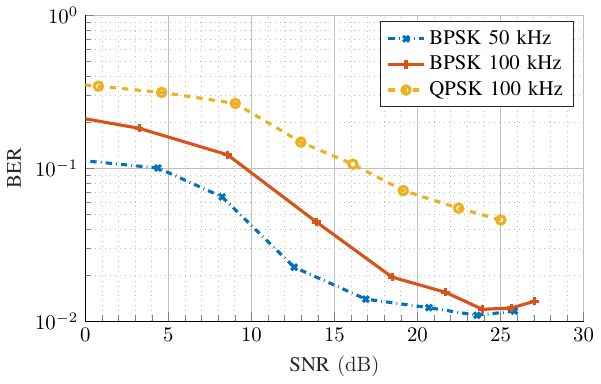}
    \caption{Bit Error Rate versus Signal-to-Noise Ratio of ZP-OFDM with differentially encoded BPSK and QPSK symbols.}
    \label{fig:ofdm_ber}
    \figsp
\end{figure}

\textbf{BER Performance Analysis.} We establish a communication link using ZP-OFDM with the setup shown in Fig.~\ref{fig:deploymentsc}. The packet structure consists of a preamble sequence for detection and synchronization and OFDM symbols separated by guard intervals. The symbols consist of $8192$ subcarriers \cite{stojanovic06oceans,tadayon19joe}. We explore a \SI{50}{\kilo\hertz} bandwidth with a BPSK modulation, and \SI{100}{\kilo\hertz} bandwidth with BPSK and QPSK modulations. The data rates of the resulting configurations are $28.8$ kbit/s, $57.6$ kbit/s, and $115.2$ kbit/s, respectively. 
We attenuate the transmission levels by $2.5$~dB steps to adjust the received SNR. We obtain the BER vs SNR results that are illustrated in Fig.~\ref{fig:ofdm_ber}. Lowering bandwidth and reducing the order of modulation results in improved BER. However, increasing bandwidth from $50$~kHz to $100$~kHz, while raising the BER, effectively doubles the data rate. Also, employing complex modulation like QPSK elevates the BER, but it too doubles the data rate compared to BPSK. This balance between BER and data rate is valuable in high-data applications like video streaming where some bit corruption is permissible.

We conducted a longer distance test of our system, utilizing the same location but moving one of the platforms approximately $160$~m away. 
To improve the transmit response levels, we implemented a transformer-based matching circuit with the transducer, which, while extending the range, limits the available system bandwidth. 
At these distances, the frequency-dependent path loss, especially at the higher frequencies used by our platform, begins to present challenges.

\begin{figure}[h]
    \centering
    \includegraphics[width=\linewidth]{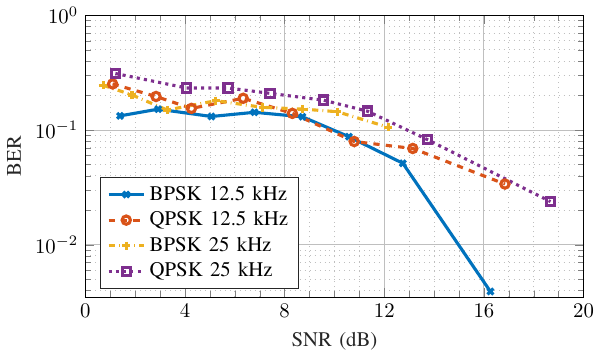}
    \caption{Bit Error Rate versus Signal-to-Noise Ratio of ZP-OFDM with differentially encoded BPSK and QPSK symbols for $160$~m distance.}
    \label{fig:ofdm_ber_long}
    \figsp
\end{figure}

As a result, we narrowed the bandwidth of our communication link to between $12.5$~kHz and $25$~kHz to manage these limitations. 
The results of BER vs.\ SNR from this test are depicted in Fig.~\ref{fig:ofdm_ber_long}. 
The data shows that system performance is affected by varying channel conditions and experiences notable fluctuations over time. 
In these tests, the peak SNR levels were lower than those in shorter-distance experiments, largely due to the limitations of the hardware, including communication modules and transducers. 
Additionally, the average results across different modulation and bandwidth settings varied, reflecting the changing instantaneous channel conditions. 
Nevertheless, the experiments demonstrate that by adjusting the physical layer parameters, satisfactory error performance can still be achieved at this range.

\subsection{Applications}

In this section we showcase the variety of use cases and applications supported by our networking platform.
First we establish a high data rate link with capable of supporting applications like underwater video streaming.
Significantly, our platform's OFDM physical layer and software-defined design achieve a data transfer rate of $153$~kbit/s, outperforming commercial modems.
This demonstrates the platform's adaptability in configuring data rates to suit application needs and reliability constraints. 
Then, we demonstrate end-to-end reliable transmission using our platform, emphasizing its utility as a network node that supports applications like file transfer.
Lastly, we demonstrate the platform's compatibility with heterogeneous networks by implementing the JANUS standard.

\subsubsection{High Data Rate Communication Link}

For applications like video streaming, a high data rate device is paramount, however, none of the commercially available platforms is able to achieve data rates over $80$~kbit/s.
As previously mentioned, the software-defined design of the SEANet networking platform can be leveraged to optimize the link to achieve high data rates. 
For the following experiment, we used broadband transducers Neptune D/140, and Teledyne Reson TC4013 on the transmitter and receiver side respectively.
The physical layer utilizes a $208.33$~kHz bandwidth on a $150$~kHz center frequency. For detection and synchronization, a maximum length sequence (m-sequence) preamble of 511 symbols is used with cross-correlation method. To reduce the packet duration and effective overhead, $10$ ms guard intervals are used along with $4$ OFDM blocks per frame. 
The OFDM blocks are loaded with differential QPSK modulated symbols over $8192$ subcarriers, with a subcarrier bandwidth of $25.4$ Hz. The payload per packet is $8064$ bytes and the raw data rate of this configuration is $307.6$~kbit/s. 
For forward error correction (FEC), we use a rate $R=1/2$ convolutional code with constraint length 12, and generator $[4335_8,5723_8]$, along with termination. This channel coding, in conjunction with a random interleaver, yields a coded data rate of $153.6$~kbit/s, nearly triple the data rate achievable by commercial modems.

\begin{figure}
    \centering
    \includegraphics[width=\linewidth]{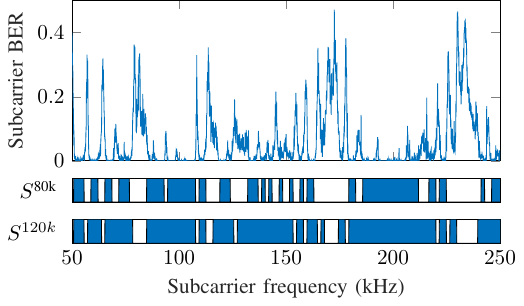}
    \caption{Measured error distribution over subcarriers and subcarrier utilization map for $80$~kbit/s and $120$~kbit/s rates.}
    \label{fig:subc_dist}
\end{figure}

By exploiting the OFDM physical layer and the software-defined architecture of the SEANet networking platform, we can optimize the use of the best-performing subcarriers, thereby balancing performance and data rate. This problem is studied in \cite{campello99,radosevic14joe} in a more general framework. We commence with the transmission of a training sequence to measure the performance, following which we adjust the PHY parameters accordingly. Packets are then generated using these parameters, and their performance is measured. We denote $K$ as the number of data subcarriers. For a given set of $N_P$ packets, each with $N_B$ blocks, we represent $e_k^{p,b}$ as the number of bit error observations in packet $p$, block $b$, on subcarrier $k$. The estimated error probability for each subcarrier $k$ is defined as per Eq.(~\ref{eq:err_prob}).
\begin{equation}
    \hat{P}_{e,k}^{\text{S}} = \frac{1}{N_P N_B} \sum^{N_P}_{p=1}\sum_{b=1}^{N_B} e_k^{p,b} , \quad \forall k\in\lbrace0,\dots,K-1\rbrace.
    \label{eq:err_prob}
\end{equation}
We transmit the training sequences with parameters $N_P=88$, $N_B=4$, $K=8064$ and estimate the $\hat{P}_{e,k}^{\text{S}}$ distribution over subcarriers as shown in the top part of Fig.~\ref{fig:subc_dist}. We observe that certain subcarriers are more error-prone, while others maintain a low error probability. To retain differential coherence among consecutive subcarriers, we divide them into $G$ groups, such that group $i$ consists of $g_i=\lbrace k_{K(i-1)/G},\dots,k_{Ki/G-1}\rbrace$. Following Eq.~(\ref{eq:err_prob}), we define the probability of error of a group $g_i$ as $\hat{P}_{e,i}^{\text{G}} = \sum_{k\in g_i} \hat{P}_{e,k}^{\text{S}}$. Finally, in Eq.~(\ref{eq:optimize}), we formulate the optimization problem that gives the minimum error rate for a given minimum data rate $R$. We define the packet duration $T_P$ and the number of data bits $N_D$, which depends on data subcarriers, modulation order, and code rate. In this formulation, each element of the vector $S=\lbrace s_1,\dots,s_G\rbrace$ indicates the usage status of a group.

\begin{equation}
\label{eq:optimize}
\begin{aligned}
		\underset{S=(s_1,\dots,s_G)}{\text{minimize}} & \quad \sum_{i=1}^{G} s_i \hat{P}_{e,i}^{\text{G}}\\
		\text{subject to} & \quad \sum_{i=1}^G s_i \geq R \frac{G\ T_P}{N_D N_B} \\
       & \quad s_i \in \lbrace0,1\rbrace
\end{aligned}
\end{equation}

\begin{figure}[h]
    \centering
    \includegraphics[width=\linewidth]{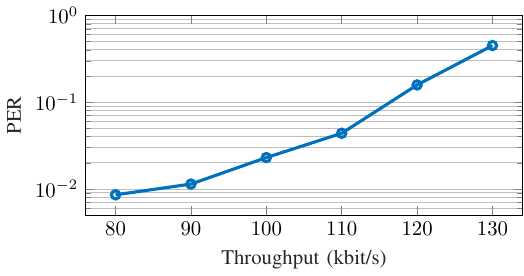}
    \caption{Packet Error Rate versus Throughput with rate adaptation based on OFDM subcarrier utilization.}
    \label{fig:per_rate}
\end{figure}

We then solve the problem for $N_D=8064$, $N_B=4$, $G=128$, $T_P=209.7$~ms and target rates $R=[80,90,\dots,130]$~kbit/s and obtain $S$ for each rate. The solutions of the optimization problem for $80$~kbit/s and $120$~kbit/s are shown in the lower part of Fig.~\ref{fig:subc_dist}. The subcarriers in groups with $s_i=0$ are treated as null subcarriers, which also increases the power loaded to the rest of the subcarriers, potentially lowering the error rate for them. We repeat the experiment by using the calculated subcarrier assignments and measure the Packet Error Rate (PER) which is shown in Fig.~\ref{fig:per_rate}. We showcase a large range of PER since for loss-tolerant applications we can use the higher data rates, for loss-intolerant applications we can choose lower data rates and lower corresponding PERs. For example, high data rates could support real-time video streaming, whereas control and telemetry streams would require a high degree of reliability as described in~\cite{campagnaro20jmse}.

\subsubsection{End-to-end Reliable Transmission}

Ensuring timely and reliable message delivery is crucial, particularly in scenarios where guaranteed delivery is essential to mitigate data loss due to channel conditions and interference~\cite{jiang18comst}. 
For instance, offshore monitoring applications generate critical sensor data around infrastructures such as underwater pipelines or offshore oil platforms~\cite{wang18oceaneng}. 
Although these systems often use wired connections for data transmission, exploiting the advantages of underwater acoustic networks necessitates the assurance of data reliability~\cite{jawhar19tii}.
Moreover, this reliability is a fundamental requirement for enabling substantial data transfers, like file transfers, from underwater nodes.

In this section, we demonstrate the potential of our proposed platform to establish reliable bidirectional links using link layer protocols. Specifically, we implement a stop-and-wait Automatic Repeat reQuest (ARQ) protocol~\cite{kurose} to establish an end-to-end reliable communication link. Briefly, the protocol involves the sender transmitting variable-sized data packets and the receiver responding with ACKs or NACKs based on packet integrity. On receipt of a NACK or if neither ACK nor NACK is received within a pre-defined time (causing a timeout), the sender retransmits the packet. The physical layer, based on chirp spread spectrum, is identical for both forward and feedback links, with bandwidths of $125$~kHz and $31.125$~kHz respectively. To simulate increased packet loss, we apply $22$~dB attenuation to the transmitter. 

\begin{figure}[h!]
    \centering
    \includegraphics[width=\linewidth]{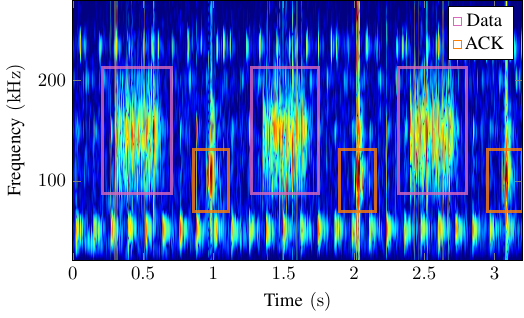}
    \caption{Recording of three packets and their ACKs in time.}
    \label{fig:arq_spectrum}
\end{figure}

\begin{figure}[h!]
    \centering
    \includegraphics[width=\linewidth]{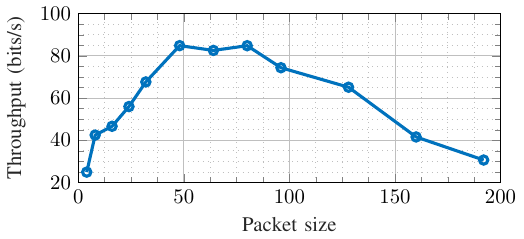}
    \caption{Throughput achieved with different packet sizes.}
    \label{fig:arq_throughput}
\end{figure}

In this set of experiments, a third platform is used for recording the spectrum activity on the channel. From Fig.~\ref{fig:arq_spectrum}, we can observe the recorded packets and their corresponding ACKs transmitted to realize the implemented ARQ mechanism. Moreover, we can also observe additional noise sources that are present in the channel such as the one spotted at around $50$~kHz. Using the implemented ARQ method, we investigate the impact of varying packet size on the communication link's throughput. As shown in Fig.~\ref{fig:arq_throughput}, we observe that the optimal throughput is achieved using packet sizes between 48 and 80 bytes in the current experimental setting. Smaller packets yield lower throughput due to the overhead of the feedback mechanism, while larger packets lead to reduced throughput due to increased PER, necessitating a high number of retransmissions.

\subsubsection{Heterogeneous Networks}
Another application for our platform is its deployment in heterogeneous networks, which comprise a variety of vehicles, sensors, and buoys. 
These network nodes present diverse communication needs and deployment constraints, which restrict compatibility across different vendors platforms. 
Our platform acts as a conduit among these systems, enhancing the network by processing and forwarding packets among other underwater assets. 
To perform these functions, our platform's software architecture is designed to emulate different waveforms and protocols. 
Furthermore, as highlighted in Section~\ref{sec:modular_design}, our platform supports a broad spectrum of frequencies. 
This feature allows it to communicate with various modems, thereby supporting interoperability between systems.

To demonstrate our proposed platform's capabilities and flexibility, we tested it with the JANUS underwater communication standard~\cite{PotterAGZNM14}, using a software-based JANUS toolkit~\cite{janusToolkit}. Despite not being specifically designed for streaming, the JANUS implementation can perform this function. As NATO's sole open-source underwater communication standard, JANUS is increasingly supported by commercial underwater modems, as discussed in Section~\ref{sec:soa}. This compatibility allows our platform to integrate with networks from other manufacturers as needed. The JANUS toolkit, which utilizes external libraries like FFTW, interacts with the waveform streamer and recorder on our system. The JANUS software operates fully on the platform, with the user connection serving merely for communication initiation and real-time data visualization.

In our demonstration, the toolkit processes I/Q samples in baseband mode with raw file format, and frequency conversion is performed on the modem. For practical purposes, we selected $100$~kHz center frequency to match our hardware specifications evaluated in Section~\ref{sec:exp_channel_response}. The resulting configuration occupies larger frequencies than typical JANUS systems operating around $10$~kHz, which can be utilized to increase the data rate with as discussed in~\cite{li23joe}.

\begin{figure}[h]
    \centering
    \includegraphics[width=\linewidth]{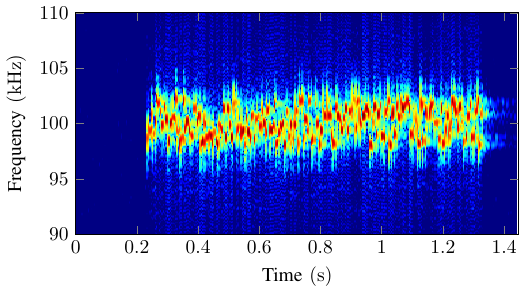}
    \caption{Spectrogram of a received JANUS packet.}
    \label{fig:janus_spectrum}
\end{figure}

\begin{figure}[h]
    \centering
    \includegraphics[trim={0 7cm 0 0},clip,width=\linewidth]{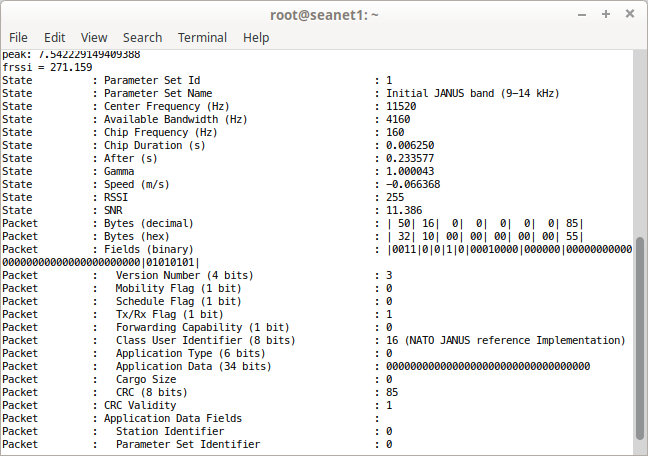}
    \caption{Output of \textit{janus-rx} of a received packet.}
    \label{fig:janus_output}
\end{figure}

The data stream transmission is achieved by interfacing file input/output of the toolkit to the named pipes connected to DMA interfaces. In addition, significant portion of the signal up/down-conversion if offloaded to the PL by reducing the receiver's sampling rate to $19.5$~ksps to decrease processing overhead and latency.
Fig.~\ref{fig:janus_spectrum} displays a received packet's spectrogram with this setup, showing a discernible frequency hopping pattern Fig.~\ref{fig:janus_output} presents the JANUS toolkit output when a packet is received.

\section{Conclusions}
\label{sec:conclusions}

The paper concerns the design, prototyping, and testing of the SEANet software-defined networking platform for underwater wireless networks. 
The platform has been engineered to address the prevalent challenges in underwater communication by integrating flexibility, high data throughput, and real-time adaptability into its core design. 
Throughout various deployments and rigorous testing in oceanic environments, SEANet has demonstrated its effectiveness by significantly outperforming existing commercial underwater modems in terms of data rate, achieving 
more than double the data rates of the fastest commercial modems.
Our experimental results have validated the platform versatility across multiple aspects, including its ability to perform reliable data transmission, establish OFDM links with different bandwidths and modulations, and integrate seamlessly with standard communication protocols like JANUS. 
Furthermore, the adaptability of SEANet is highlighted through its successful operation under different configurations and conditions, showcasing its potential to support diverse underwater operations—from data-heavy scientific research to critical infrastructure monitoring.
In conclusion, the SEANet platform stands as a significant advancement in the field of underwater communication, offering a scalable, flexible, and high-performance solution with the potential to revolutionize how we deploy and utilize the Internet of Underwater Things.

\bibliographystyle{IEEEtran}
\bibliography{references}

\clearpage

\begin{IEEEbiography}
[{\includegraphics[width=1in,height=1.25in,keepaspectratio]{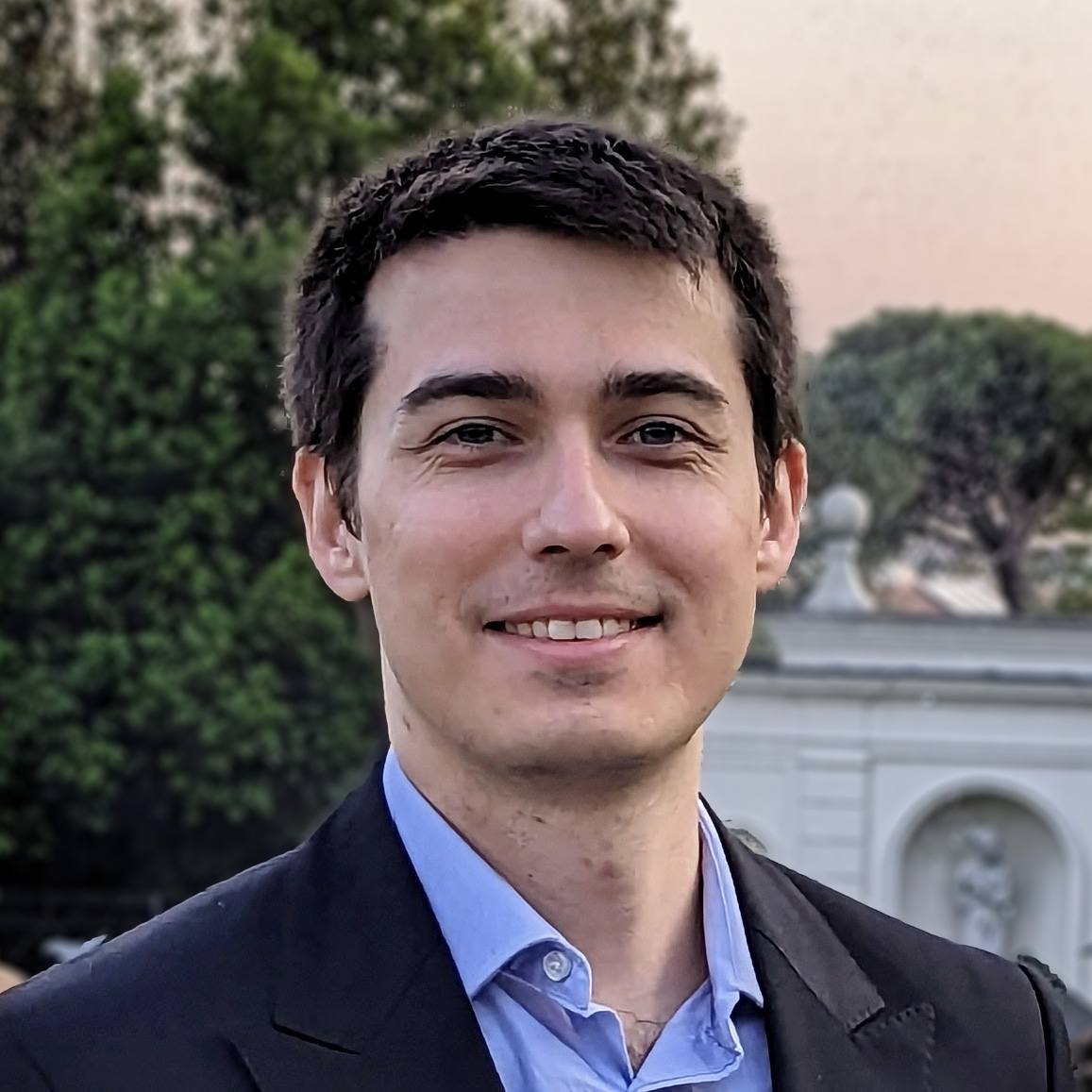}}]{Deniz Unal}
is a Ph.D. candidate in electrical engineering at the Institute for the Wireless Internet of Things and the Department of Electrical and Computer Engineering at Northeastern University. He is working as a research assistant under the supervision of Tommaso Melodia. He received his BS and MS degrees in electrical and electronics engineering from Bilkent University, Ankara, Turkey, in 2014 and 2018, respectively. His research interests include underwater wireless networks, underwater acoustic communications, and software-defined networking.
\end{IEEEbiography}

\begin{IEEEbiography}
[{\includegraphics[width=1in,height=1.25in,keepaspectratio]{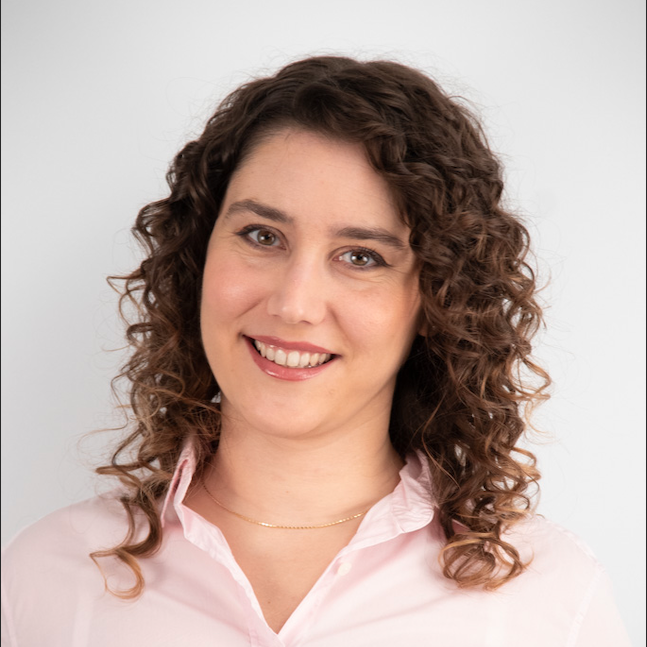}}]{Sara Falleni}
is a Ph.D. candidate in Computer Engineering at the Department of Electrical and Computer Engineering and the Institute for the Wireless Internet of Things at Northeastern University, Boston, MA, USA. She is currently working with Prof. Stefano Basagni and her research interests include the adaptability and reliability of high data rate underwater acoustic wireless networks. She received the B.Sc. degree in Computer Engineering from the University of Pisa, Pisa, Italy, in 2014, and the M.S. degree (Hons.) in Embedded Computing Systems from the Scuola Superiore Sant’Anna of Pisa, Pisa, Italy, in 2017.
\end{IEEEbiography}

\begin{IEEEbiography}
[{\includegraphics[width=1in,height=1.25in,keepaspectratio]{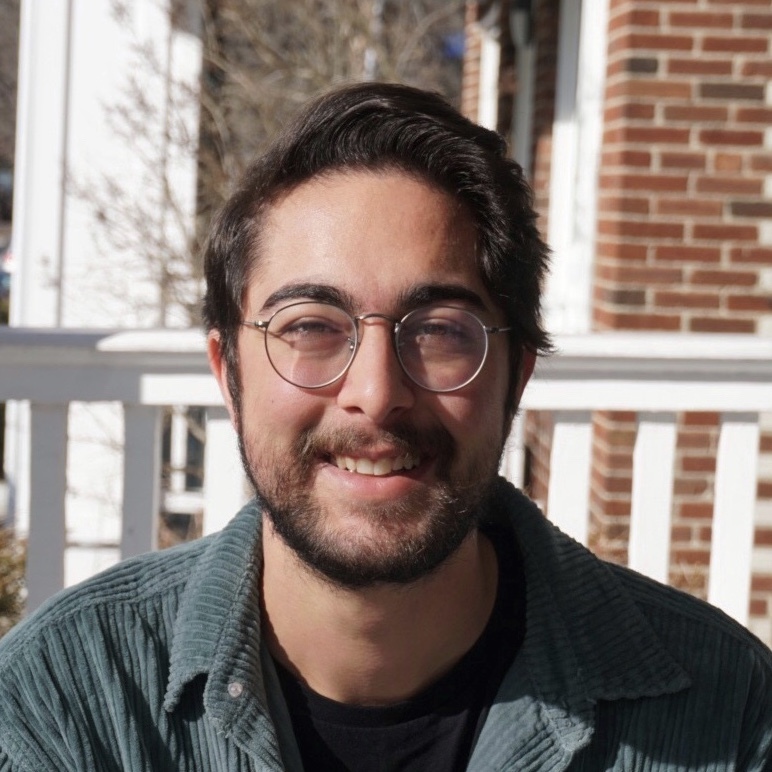}}]{Kerem Enhos}
received the B.S. and M.S. degrees in electrical and electronics engineering from Bilkent University, Ankara, Turkey, in 2017 and 2019, respectively, and the Ph.D. degree from Northeastern University, in 2023, all in electrical engineering. He is currently Senior Research and Development Engineer with Hydronet. His research interests include underwater wireless communications, networking, and systems; acoustic waves and devices; visible light communications and software-defined networking.
\end{IEEEbiography}

\begin{IEEEbiography}
[{\includegraphics[width=1in,height=1.25in,keepaspectratio]{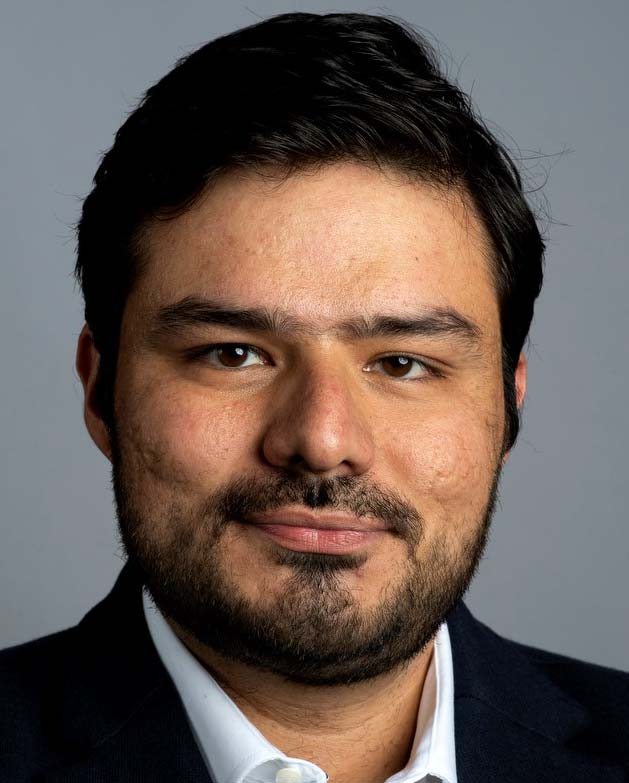}}]{Emrecan Demirors}
received the Ph.D. degree in electrical and computer engineering from Northeastern University in 2017. He is currently a Research Assistant Professor with the Department of Electrical and Computer Engineering, Northeastern University, where he is also a member of the Institute for the Wireless Internet of Things. His current research interests include the Internet of Things and 5G networks, underwater communications and networks, Internet of Medical Things, unmanned aerial, and underwater vehicle networking. He is an Associate Editor of IEEE Access. He organized the IEEE International Workshop on Wireless Communications and Networking in Extreme Environments from 2017 to 2022. He has also been serving as a TPC Member for IEEE WCNC since 2018, IEEE PIMRC since 2020, and among others. He is the Co-Founder and Director of R\&D of Bionet Sonar.
\end{IEEEbiography}

\begin{IEEEbiography}
[{\includegraphics[width=1in,height=1.25in,keepaspectratio]{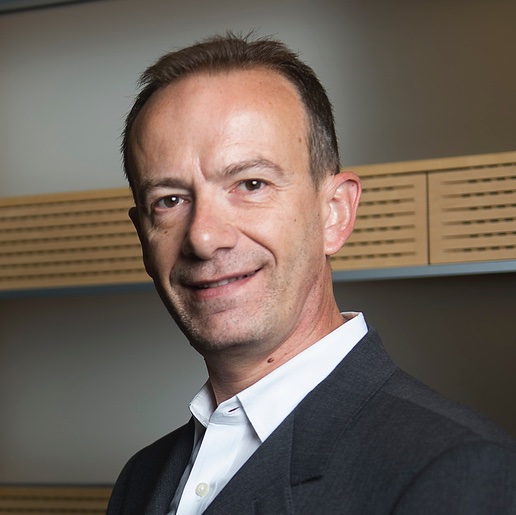}}]{Stefano Basagni}
is with the Institute for the Wireless Internet of Things and a professor at the ECE Department at Northeastern University, in Boston, MA. He holds a Ph.D. in electrical engineering from the University of Texas at Dallas (2001) and a Ph.D. in computer science from the University of Milano, Italy (1998). Dr. Basagni’s current interests concern research and implementation aspects of mobile networks and wireless communications systems, wireless sensor networking for IoT (underwater, aerial and terrestrial), and definition and performance evaluation of network protocols. Dr. Basagni has published over twelve dozen of highly cited, refereed technical papers and book chapters. His h-index is currently 49 (November 2022). He is also co-editor of three books. Dr. Basagni served as a guest editor of multiple international ACM/IEEE, Wiley and Elsevier journals. He has been the TPC co-chair of international conferences. He is a distinguished scientist of the ACM, a senior member of the IEEE, and a member of CUR (Council for Undergraduate Education).
\end{IEEEbiography}

\begin{IEEEbiography}
[{\includegraphics[width=1in,height=1.25in,keepaspectratio]{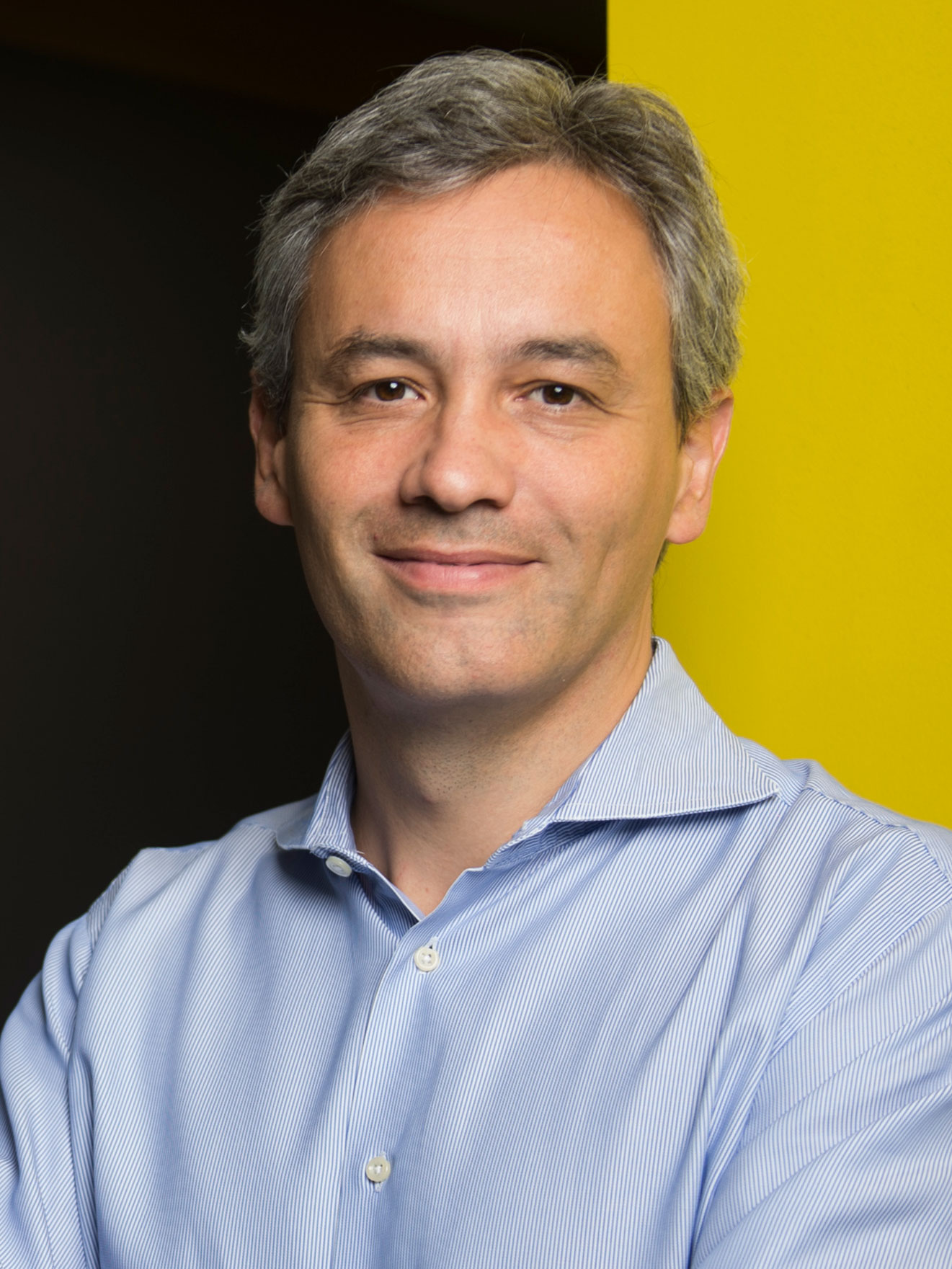}}]{Tommaso Melodia}
is the William Lincoln Smith Chair Professor with the Department of Electrical and Computer Engineering at Northeastern University in Boston. He is also the Founding Director of the Institute for the Wireless Internet of Things and the Director of Research for the PAWR Project Office. He received his Ph.D. in Electrical and Computer Engineering from the Georgia Institute of Technology in 2007. He is a recipient of the National Science Foundation CAREER award. Prof. Melodia has served as Associate Editor of IEEE Transactions on Wireless Communications, IEEE Transactions on Mobile Computing, Elsevier Computer Networks, among others. He has served as Technical Program Committee Chair for IEEE Infocom 2018, General Chair for IEEE SECON 2019, ACM Nanocom 2019, and ACM WUWnet 2014. Prof. Melodia is the Director of Research for the Platforms for Advanced Wireless Research (PAWR) Project Office, a \$100M public–private partnership to establish 4 city-scale platforms for wireless research to advance the US wireless ecosystem in years to come. Prof. Melodia’s research on modeling, optimization, and experimental evaluation of Internet-of-Things and wireless networked systems has been funded by the National Science Foundation, the Air Force Research Laboratory the Office of Naval Research, DARPA, and the Army Research Laboratory. Prof. Melodia is a Fellow of the IEEE and a Senior Member of the ACM.
\end{IEEEbiography}

\end{document}